\documentclass{ws-procs9x6}
\def\be{\begin{eqnarray} &&}

\def\ee{\end{eqnarray}}
\def\psla{\rlap \slash}

\usepackage{graphicx} 
\usepackage{dcolumn} 
\usepackage{bm} 
\usepackage{epsfig}

\begin{document}

\title{POLE APPROXIMATION FOR PION ELECTROMAGNETIC FORM FACTOR
WITHIN LIGHT-FRONT DYNAMICS}

\author{J. P. B. C. de Melo$^{a,b}$ and J. S. Veiga$^a$}
 \address{$^a$Centro de Ci\^encias Exatas e Tecnol\'ogicas,
Universidade Cruzeiro do Sul, 08060-070, S\~ao Paulo, SP, Brazil \\
$^b$Instituto de F\'\i sica Te\'{o}rica, Universidade Estadual
Paulista, 01405-900, S\~{a}o Paulo, SP, Brazil}
\author{T. Frederico}
\address{Departamento de F\'isica, Instituto Tecnol\'ogico de Aeron\'autica,   \\
12228-900, S\~{a}o Jos\'e dos Campos, SP, Brazil}
\author{E. Pace}
\address{Dipartimento di Fisica,
Universit\`a di Roma "Tor Vergata" and Istituto
Nazionale di Fisica Nucleare, Sezione Tor Vergata, Via della Ricerca
Scientifica 1, I-00133, Roma, Italy}
\author{G. Salm\'e}
\address{Istituto Nazionale di Fisica Nucleare,
Sezione Roma I, P.le A. Moro 2,
 I-00185,  Roma, Italy }

\begin{abstract}
We carefully investigate the reliability of the propagator Pole
Approximation, i.e. the approximation of retaining only the
propagator poles in the evaluation of the Mandelstam covariant
expression for the electromagnetic current of the pion. Different
frames are analyzed, in order to find the most suitable one for
calculating the pion form factor within the proposed
approximation. It turns out that the approximation is more
accurate in the frame where $q^+$ is maximal. The relevance of the
Pole Approximation is briefly discussed in view of calculations of
hadron form factors based on wave functions generated by dynamical
models.\end{abstract}

\section{Introduction}

Light-front field theory  starts with the paper by Dirac in 1949,
when he proposed different forms to describe relativistic
systems~\cite{Dirac49}. After Dirac, Light-front field theory  was
applied to calculated many process in relativistic nuclear physics
and particle physics (see the reviews by Brodsky, Pinsky,
Pauli~\cite{Brodsky98} and Harindranath~\cite{Harina96} for
details). In parallel, the
description of the electromagnetic properties of hadronic systems
within the light-front dynamics framework, based on the introduction
of a valence bound state wave function, has found some
success~
\cite{Terentev76,Dziembowski87,Cardarelli95,Pacheco97,Pacheco99,Pacheco992}.
In the light-front dynamics  the bound state wave
functions are defined on the hyper-surface $x^+=x^0+x^3=0$. These
wave functions are covariant under kinematical front-form boosts,
due to the stability of Fock-state decomposition under these
boosts~\cite{Perry90,Wilson95}.

In recent works, (see, e.g., \refcite{Pacheco2004,Pacheco20052}), a third path
for evaluating hadron electromagnetic (em) form factors has been proposed.
The Mandelstam covariant expression for the matrix elements of the current has
been calculated replacing the  quark-hadron vertex function in the valence
range
by a hadron wave function obtained in a dynamical model,  able to
reproduce the energy spectrum. In those works the analytical structure of
quark-hadron vertex function has been neglected, retaining only the poles from
the constituent propagators.
In this contribution we will study the reliability of such an approximation
within an analytic, covariant model for the pion em
form factor~\cite{Pacheco2002}.

\section{Electromagnetic Pion Form Factor}

In general, the pion em form factor  is
given by
\begin{equation}
P^\mu F_{\pi}(q^2)\ =
\langle\pi(p^{\prime})|J^{\mu}|\pi(p)\rangle,
\end{equation}
where $P=p+p^{\prime}$, $q=p^\prime -p$ and $J^{\mu}$ is the
electromagnetic current operator. It is expressed in terms of the
quark fields~$q_f$ and charge $e_f$~($f$ is the flavor of the quark
field): $J^{\mu}=\sum_{f} e_{f} \bar{q}_{f} \gamma_\mu q_{f}$.

In the impulse approximation the em form factor is
given by:
\begin{eqnarray}
P^\mu F_\pi(q^2)= \int \frac{d^4k}{(2 \pi)^4} Tr\Bigr[ \gamma^5
S(k-p^{\prime}) \gamma^\mu S(k-p) \gamma^5 S(k) \Bigr]
\Gamma(k,p^{\prime}) \Gamma(k,p) \label{ffactor0}
\end{eqnarray}
where   $\Gamma(k,p)$ is the pion-quark vertex function.  Here, we
will show calculations for a vertex function with a symmetric
form, previously used in Ref.~\refcite{Pacheco2002}:
\begin{equation}
\Gamma(k,p)= \biggl[ \frac{N}{(k^2-m^2_R + \imath\epsilon)} +
\frac{N}{((p-k)^2-m^2_R + \imath\epsilon)} \biggr]. \label{symm}
\end{equation}

The pion em form factor is extracted from the
$J^{+}_{\pi}$ current using Eq.~(\ref{ffactor0}):
\begin{eqnarray}
F_\pi(q^2) & = &
 \frac{N_c~m^2}{f^2_\pi  (p^{+}+p^{\prime +}) } \int
\frac{dk^-dk^+d^2k_{\perp}} { 2 (2\pi)^4} \frac{
\Gamma(k,p^{\prime}) \Gamma(k,p)} {k^+ (k^- - \frac{f_1-\imath
\epsilon}{k^+})} \\ \times & & Tr\biggl[ \frac{ (\psla{k}+m)
\gamma^5 (\psla{k}-\psla{p^{\prime}}+m) \gamma^{+}
(\psla{k}-\psla{p}+m) \gamma^5 ] }{(p^+ - k^+)(p^{-}-k^- -
\frac{f_2 -\imath \epsilon }{p^{+} - k^+}) (p^{\prime+} -
k^+)(p^{\prime-}-k^- - \frac{f_3-\imath \epsilon } {(p^{\prime+} -
k^+) } } \label{ffactor} \biggr], \nonumber
\end{eqnarray}
where, $f_1=k_{\perp}^2+m^2$, $f_2=(p-k)_{\perp}^2+m^2$ and
$f_3=(p^{\prime}-k)_{\perp}^2+m^2$.  The $k^+$
momentum integration has two contributions: (i) $0<k^+<p^+$ and
(ii) $p^+<k^+<p^{\prime +}$, where $p^{\prime +}=p^+ + q^+$. The
first interval (i) in the $k^+$ integration includes the
contribution of the valence component of the wave function and the
second one (ii) corresponds to the contribution of the pair
term~\cite{Pacheco98}.

The pion valence wave function for the symmetric vertex is given
by~\cite{Pacheco2002}:
\begin{eqnarray*}
\Psi(k^+,\vec k_\perp) & = & \left[\frac{{ N}}
{(1-x)(m^2_{\pi}-{ M}^2(m^2, m_R^2))} \right. \left.
+\frac{{ N}} {x(m^2_{\pi}-{ M}^2(m^2_R, m^2))}
\right]  \\
& & \times \frac{p^+}{m^2_\pi-M^2_{0}} .
\label{wf2}
\end{eqnarray*}
where $x=k^{+}/p^{+}$ is the fraction of the momentum carried  by the quark.
 The function ${ M}$  is defined as
 \begin{equation}
{ M}^2(m^2,m^2_R)= \frac{k_{\perp}^2+m^2}{x}+
\frac{(p-k)_{\perp}^2+m_R^2}{(1-x)}-p^2_{\perp} \ .
\end{equation}
The free square mass of the quark-antiquark system is ${
M}^2_0={ M}^2(m^2,m^2)$  and the normalization constant $N$ is
found from the condition $F_{\pi}(0)=1$.

The parameters of the model are the quark mass $m_{q}=0.220$~GeV,
the regulator $m_{R}=0.60$~GeV mass and the experimental pion mass
$m_{\pi}= 0.140$~GeV~\cite{Pacheco2002}. Our choice of the
regulator mass value fits the pion decay constant
$f_{\pi}^{exp}=92.4$~MeV.

\section{Propagator Pole Approximation}

Taking into account only the poles coming from the
quark propagators in Eq. (\ref{ffactor}), the pion
em form factor is calculated from: i) the residue of
the pole of the spectator quark on-minus-shell ({\em valence} contribution), and
ii) a
contribution coming from  the pair production mechanism
({\em non-valence} contribution). The pole approximation~($PA$) is
compared with the full result in different reference frames, labeled by
different values of the plus component of the momentum transfer.
In order to accomplish the freedom of changing the frames, let us parametrize
the momentum transfer as follows:  $q^+=-q^-=\sqrt{-q^2} \sin\alpha$,
 $q_x=\sqrt{-q^2}\cos\alpha$
and $q_y=0$.

\begin{figure}[thb]
\centering
\includegraphics[width=3.0 in,angle=-90]{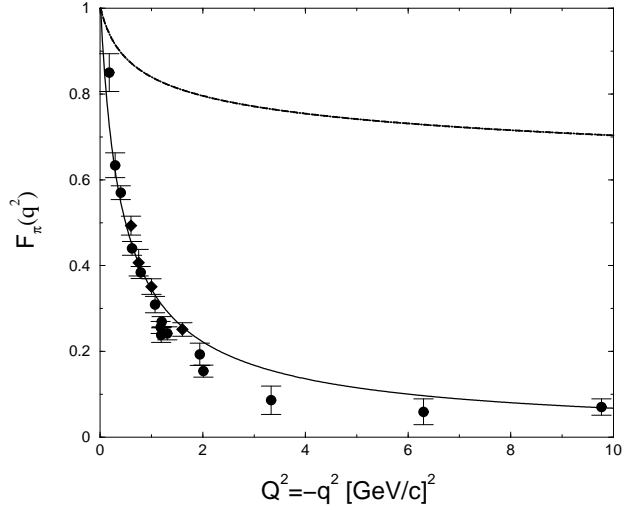}
\caption{Pion electromagnetic form factor vs square momentum
transfer  $Q^2=-q^2$ for $\alpha=0$. Lower solid curve: full
calculation. Upper solid curve: pion Pole Approximation.
Experimental data from the compilation of Baldini et al. [18]}
\end{figure}

The poles in the $k^-$ integration contributing to the pion
em current in the triangle diagram from the quark
propagators,  in the PA, correspond to two
intervals in the $k^+$ integration: the first is $(i) ~0<k^+ <
p^+$ and in this case, the pole is ~$k^{-}_1 =
(k_{\perp}^2+m^2/{k^+}$ (valence region) and, the second one
is (ii) $p^+ < k^+ <p^{\prime+}$, and we consider the pole
$k^{-}_3=p^{-} - (p^{\prime}-k)_{\perp}^2+m^2)/ {(p^{\prime
+}-k^+)}$ (pair term region).

The em pion form factor within the PA
is given by
\begin{equation}
F^{(PA)}_{\pi}(q^2)= F^{(I)}_{\pi}(q^2)|_{k_1^-} +
F^{(II)}_{\pi}(q^2)|_{k_3^-}.
\label{PA}
\end{equation}
In the PA, the contributions coming from the
poles of the quark-pion vertex are neglected.  We have used three
values of $\alpha=0^o$ (Drell-Yan condition, $q^+=0$), $45^o$ and $90^o$.

In particular, the em form factor in
the Drell-Yan frame reduces to the valence contribution, namely
 \begin{eqnarray}
F^{(PA)}_\pi(q^2) = F^{(I)}_{\pi}(q^2)|_{k_1^-} & =  & \frac{m^2 N_c}
{p^+f_{\pi}^2} \int \frac{ d^{2}
k_{\perp}}{2 (2\pi)^3 } \int_0^{1} \- \- dx \left[ k_{on}^- p^{+
2} + \frac14 x p^{+} q^2 \right] \\ \nonumber & & \times \frac{
\Psi^{*}_{f}(x,k_\perp) \Psi_{i}(x,k_\perp)} {x (1-x)^2},
\label{ffwf}\end{eqnarray}
 where $k_{on}^-=k_{\perp}^2+m^2/k^+$.

In Figs. 1, 2 and 3, the pion form factor calculated
within the PA for $\alpha=0$, $45^o$ and $90^o$,
respectively, is presented. Results obtained
 considering both the valence contribution only
($F^{(I)}_{\pi}(q^2)|_{k_1^-}$) and
 valence plus nonvalence contributions, Eq. (\ref{PA}), are shown. The
comparison with the exact result, Eq. (\ref{ffactor}), indicates that
 in the Drell-Yan frame PA works very badly, while in the frame where
the momentum transfer has no transverse components  at all PA
appears remarkably effective, in particular at high momentum transfer.
In this kinematical region, the pair
production mechanism dominates, that can be related to the absorption of a
$q\bar q$ pair by a single constituent, that has a given longitudinal-momentum
distribution.

\begin{figure}[thb]
\centering
\includegraphics[width=3.0 in,angle=-360]{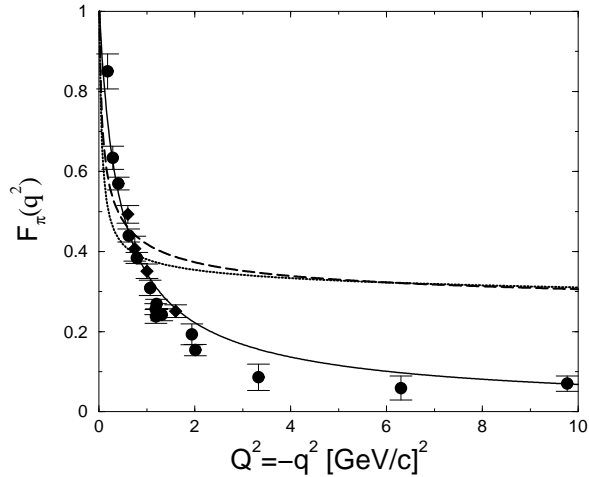}
\caption{Pion electromagnetic form factor vs
$Q^2=-q^2$ for $\alpha=45^o$. Lower solid curve: full calculation.
 Upper solid curve: valence contribution in
Pole Approximation. Dashed curve: valence + nonvalence contribution in
Pole Approximation.
 Experimental data from the compilation of Baldini
et al. [18]}
\end{figure}
\begin{figure}[thb]
\centering
\includegraphics[width=3.0 in,angle=-360]{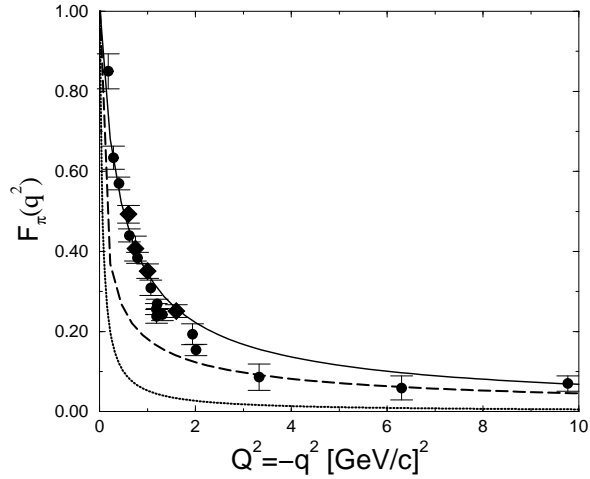}
\caption{Pion electromagnetic form factor vs
$Q^2=-q^2$ for $\alpha=90^o$.  Upper solid curve: full calculation.
 Lower solid curve: valence contribution in
Pole Approximation. Dashed curve: valence + nonvalence contribution in
Pole Approximation.
 Experimental data from the compilation of Baldini
et al. [18]}
\end{figure}

\section{Conclusions}

The pion em form factor is calculated
 for $q^2$ up to $10~[GeV/c]^2$ with the
covariant symmetric model for the quark-pion
vertex~\cite{Pacheco2002}. The exact calculation is compared with
an approximate evaluation of the form factor in frames with different values of
$q^+$. In the propagator Pole Approximation, investigated in this contribution,
 only the poles originated by the quark propagators are taken into account,
while  the poles of the quark-hadron vertex are disregarded in the analytical
integration over the light-front energy.
 We found that in the frame where $q^+$ is maximal and
the pair term dominates, the approximation is able do describe
qualitatively well the exact results.

\section*{Acknowledgments}

This work was partially supported by Funda\c{c}\~ao de Amparo \`a
Pesquisa do Estado de S\~ao Paulo, Conselho Nacional de Pesquisas
and by Ministero dell'Istruzione, dell'Universit\`a e della
Ricerca.

\end{document}